# Evidence against a general positive eddy feedback in atmospheric blocking

**Lei Wang[1,2] and Zhiming Kuang[1,2]**

[1] Department of Earth and Planetary Sciences, Harvard University, Cambridge, MA, USA.

[2] John A. Paulson School of Engineering and Applied Sciences, Harvard University, Cambridge, MA, USA.

Corresponding author: Lei Wang (leiwang@g.harvard.edu)

**Key Points:**

- Atmospheric blocks can be simulated in a zonally symmetric two-layer quasi-geostrophic model without topography.

- Randomly selected synoptic eddies do not reinforce the blocks through a positive feedback, challenging the eddy straining mechanism.

- The second-order induced flow, previously used to support the general eddy feedback idea, is sensitive to the location of the wave maker.

**Abstract**

The eddy straining mechanism of Shutts (1983; S83) has long been considered a main process for explaining the maintenance of atmospheric blocking. As hypothesized in S83, incoming synoptic eddies experience a meridional straining effect when approaching a split jetstream, and as a result, enhanced PV fluxes reinforce the block. A two-layer QG model is adopted here as a minimal model to conduct mechanism-denial experiments. While transient eddies' forcing is clearly critical to the formation and maintenance of a block, using a large ensemble, the authors demonstrate that the straining of generic eddies does not maintain blocks, thus challenge the idea of eddy straining serving as a positive feedback for the blocks. These results indicate that specific configurations of the eddy field are required for the maintenance stage. The authors also remark on the main supporting evidence in S83: the second-order induced flow is sensitive to the location of the wavemaker.

**Plain Language Summary**

Atmospheric blocking is an important process for both weather and climate. The eddy straining mechanism of Shutts (1983) has been considered as the foundation to understanding the maintenance of blocks, which is consistent with the observation that strong wave breaking is always concurrent with strong blocks. However, the authors here use a large-ensemble of a two-layer quasi-geostrophic model to assess whether such a correlation is due to a causal relationship between the straining of generic eddies by the block and block enhancement, as proposed by Shutts (1983). Surprisingly, the authors demonstrate that the generic effects from straining eddies is insignificant to the maintenance of the blocking pattern. The authors also note an important issue with the experimental design and the physical relevance of the weakly nonlinear simulations in Shutts (1983). Therefore, the authors challenge the existence of a generic mechanism of positive eddy feedback, and attribute the blocks' actual maintenance to the specific initial condition of pre-existing eddies, which intrinsically dependents on chance.

# 1 Introduction

Atmospheric blocking is an important phenomenon characterized by a quasi-stationary and large-scale meridional dipole flow fields in the mid-latitude atmosphere. A common feature across most blocking types is a meridional dipole in the anomaly field of PV or geopotential height. Once a block has been established, such a dipole structure can persist for more than the synoptic timescale, exerting impacts on local weather with significant societal implications.

During a blocking episode, synoptic weather systems propagate from upstream and can interact with the blocking dipole (Berggren et al., 1949). Such transient waves' forcing is critical for the maintenance of a block (Holopainen & Fortelius, 1987; Woollings et al., 2018). This paper is focused on the role of eddies during the maintenance stage of blocks, by addressing an open question regarding whether generic incoming eddies can reinforce a block as a positive feedback.

A seminal paper by Shutts (1983) (S83) proposed a mechanism that, as synoptic eddies approach a block, they will be strained meridionally, and the enhanced diffusion near the point of jet splitting can drive a meridionally asymmetric second-order flow. Importantly, using a weakly nonlinear model, S83 demonstrates that the numerical solution of the second-order flow has an in-phase structure that reinforces the existing block's amplitude. At the same time, observational evidence suggests that, during a mature block, warm air can move poleward and eventually cut off anticyclonically from incoming waves to directly merge with the blocking anticyclones, and the opposite is true for cold air to merge with the blocking cyclones. This is termed the selective absorption mechanism (Yamazaki & Itoh, 2012) (YI12) and the essential physical process is schematically described in Section 17.5 of a textbook by Hoskins and James (2014) (HJ14). Both mechanisms are schematically summarized in Fig. 1a-f. Note that both mechanisms assume the initial incoming eddies are meridionally symmetric, which is consistent with the observations of mid-latitude eddies. In this paper, we will not address studies using the modon solution (e.g. Haines & Marshall, 1987) because of the strong instability of a modon and the sensitivity to domain configurations (Arai & Mukougawa, 2002).

The observational evidence is indeed compelling that synoptic eddies strain and shift meridionally during a mature block. However, such evidence does not necessarily imply that any incoming eddies can systematically reinforce the block as a generic positive feedback mechanism, as hypothesized in S83, which imposed no restrictions on the incoming eddies. An alternative interpretation to a generic positive feedback would be that only under certain conditions of the pre-existing synoptic eddies can a dipole structure in eddy forcing be developed through nonlinear eddy-eddy interactions to reinforce the block. In this case, eddy straining, meridional shifts, and wave breaking would occur but there is no generic positive feedback; whether a block persists is determined by conditions of the pre-existing synoptic eddies.

To rule out the sensitivity to the initial conditions of pre-existing synoptic eddies, we aim to test whether any substantial reinforcing effect is present and identifiable when we randomly sample initial conditions from equilibrated baroclinic eddy fields. We adopt a minimal model for tackling this issue: a two-layer quasi-geostrophic model, which is fully nonlinear and baroclinically unstable without the need of a wavemaker. In the next section, we will illustrate that blocking maintenance cannot be supported by randomly selected equilibrated flow field, casting doubt on the existence of the positive eddy feedback hypothesized in S83.

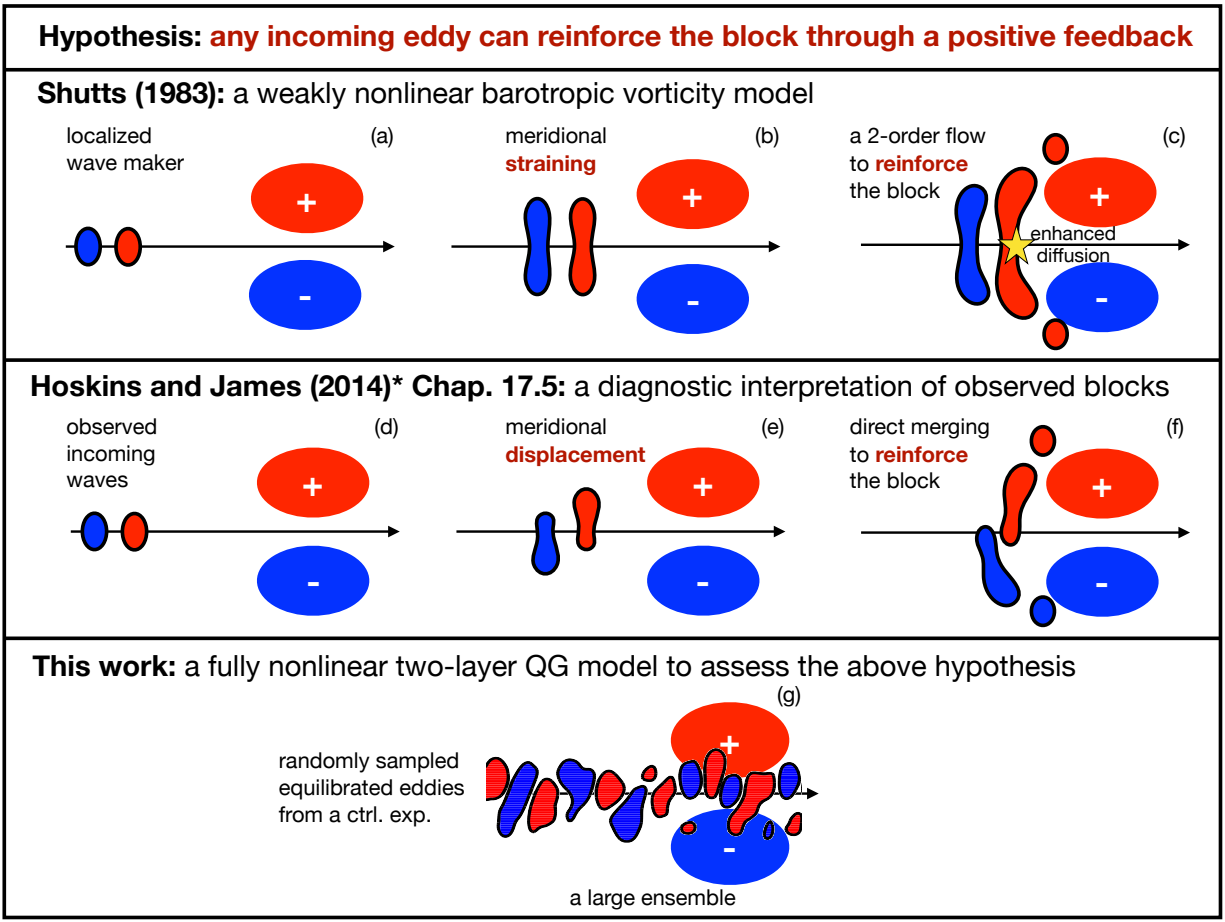

**Figure 1**. A schematic diagram of three different approaches for testing the hypothesis whether generic eddies can reinforce the block as a positive feedback.

## 2 A Mechanism-Denial Study

### 2.1 A two-layer quasi-geostrophic model

A main issue with the barotropic model used in prior studies is that incoming waves can only be prescribed with a wavemaker. A two-layer QG model is a minimal model that can intrinsically generate synoptic eddies that are equilibrated across the domain. A two-layer model has been used to study blocks (Vautard et al., 1988; K. Haines & Holland, 1998; Luo, 2000). To extract the full pattern of a mature block, we integrate a two-layer QG model for 100000 days and take the last 90000 days for composite analysis. See Appendix A for more details on the model.

We define blocking based on local finite-amplitude wave activity (LWA) (Huang & Nakamura, 2016), which bears a clear physical meaning: large values of wave activity straightforwardly correspond to an overturn of PV contours in both the north and south sides of a certain latitude. With a threshold of 1.5 standard deviation of LWA variability, 351 blocks that last at least 10 days are identified. These blocks will be used for the composite. The exactly number of blocks would of course vary with different threshold values, but salient features of the blocks remain the same for strong and persistent blocks. Fig. 2 demonstrates the evolution of total PV fields in the upper layer of the QG model from the composite.

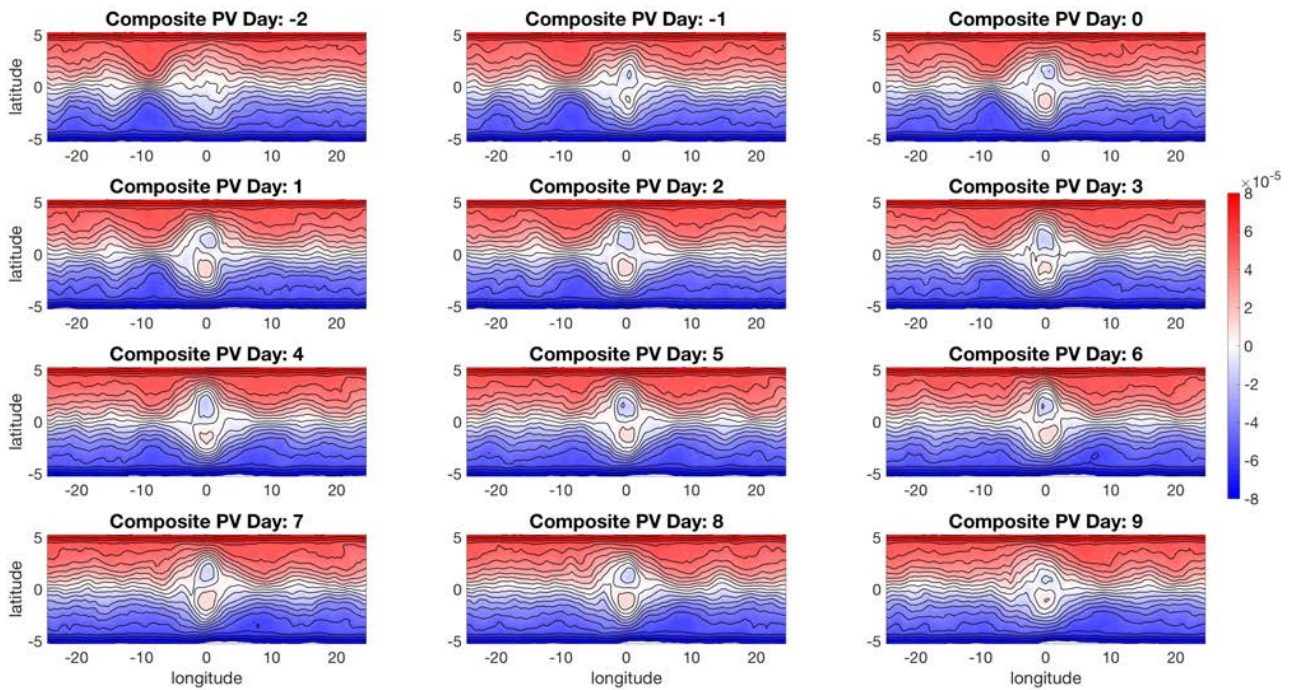

**Figure 2**. Composite blocks in a two-layer QG model. Evolution of the upper layer PV from two days prior to the onset of the block to day 9. The contour interval is $0.2\times10^{-5}$ s$^{-1}$.

**2.2 Evolution of the eddy PV flux divergence**

The ensemble-mean evolution of the barotropic eddy PV can be described as:

$$\frac{\partial [q_b']}{\partial t} + J([\psi_b],[q_b]) = -\nabla \cdot [\mathbf{v}_b' q_b'] + [D] \tag{1}$$

where $[\cdot]$ denotes ensemble average, subscript $b$ denotes barotropic component, and $D$ represents dissipation (associated with 6$^{th}$-order hyper-viscosity) and Ekman damping. During the mature stage of blocks in this model, on the leading order, the mean flow advection of block is balanced by the planetary vorticity advection and self-advection by the block (not shown), i.e. $J([\psi_b],[q_b]) \sim 0$, therefore the eddy PV flux divergence $-\nabla \cdot [\mathbf{v}'q']$ (hereafter as eddy forcing) is the dominant driving force for the ensemble-mean evolution of the eddy PV. As shown in Fig. 3, the ensemble-mean eddy forcing shows a consistent dipole structure at the beginning stage of the composite (even two days before blocking onsets), and gradually switches to a negative phase after day 8 which would eventually damp the block toward the end of the block lifecycle. Such evolution suggests that transient eddies are responsible for the full evolution of the block.

However, the importance of transient eddies does not necessarily imply the existence of a general positive eddy feedback as a result of either straining (as described by S83) or meridional displacements (as described by YI12 and HJ14) of a generic eddy field. It remains possible that, by compositing on strong and persistent blocks, we have selected eddy fields with specific configurations and those specific configurations, instead of a general eddy feedback, are the key to the resulting eddy forcing that maintains the block. Next we turn to simulations using randomly selected eddy fields in order to evaluate the existence of a general eddy feedback.

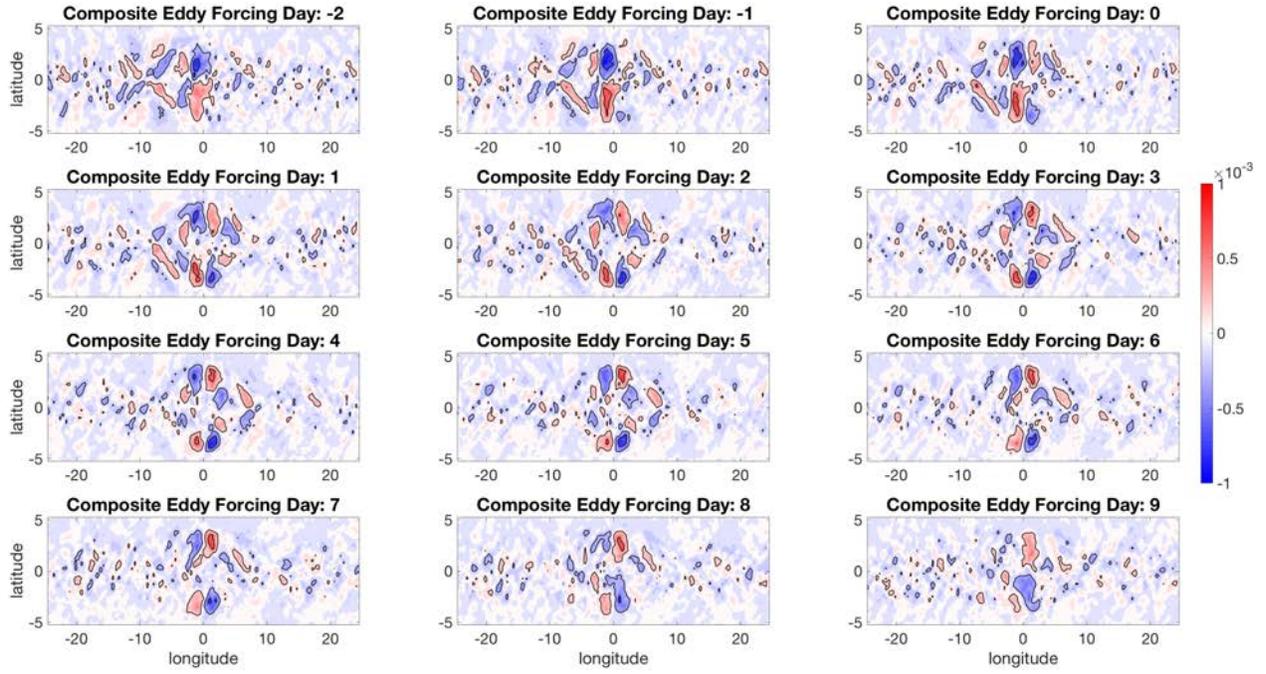

**Figure 3**. Composite Eddy PV flux divergence $-\nabla \cdot [\mathbf{v}'q']$ in a two-layer QG model, corresponding to the PV evolution in Fig. 2. The contour interval is $4\times10^{-4}$ s$^{-2}$.

**2.3 A large-ensemble simulations to assess the hypothesis**

To answer the question proposed in the last section, we now assess whether a randomly selected PV field can reinforce a block as predicted in S83. Based on the composite, we take the blocking pattern $q'_{block}$, which is defined as the anomaly field (departure from zonal-mean) at day 7 of the composite block, and add it to three thousands randomly selected PV fields $q_i$ from the control run for both layers, such that the ensemble mean of the flow field satisfies:

$$[q_i + q'_{block}] = \bar{q} + q'_{block} \qquad (2)$$

where $\bar{q}$ refers to the climatological basic state of the control run. In other words, the procedure guarantees that the ensemble-mean basic state equals to the model control run's basic state plus the mature blocking anomaly.

Under this experiment design, we stress that, in the ensemble-mean sense, while the contribution from deformed eddies (i.e. strained eddies) is well represented as it is ubiquitous in a turbulence flow, no contribution is allowed from the specific pre-existing eddies associated with blocking, because the random selection of the PV fields as initial conditions does not favor any specific configurations of such eddies.

The resulting ensemble-mean block retains its dipole structure for only 3 days (See Figure S1). This is an indication that the transient eddies, when they are randomly sampled, do not contribute substantially to the maintenance of the block. However, it is still possible that such a lack of reinforcement could due to the rapid (~3 days) collapse of the block. Therefore, we add a time-invariant corrector of PV in both layers to ensure that a mature block pattern is robustly maintained throughout the calculation. The time-invariant corrector is based on the negative tendency of the ensemble-mean drift in day 1. With this corrector, the block structure is well retained up to day 12 (See Figure S2).

A striking result is that, in Fig. 4, the ensemble-mean eddy forcing is almost zero everywhere. If the transient eddies are reinforcing the block as a positive feedback, we would expect the results, at least its order of magnitude, to be consistent with that in Fig. 3 from the composite. This result suggests that it is the specific initial condition that leads to the block maintenance, instead of a positive feedback as a result of eddy straining and meridional shifts by the block.

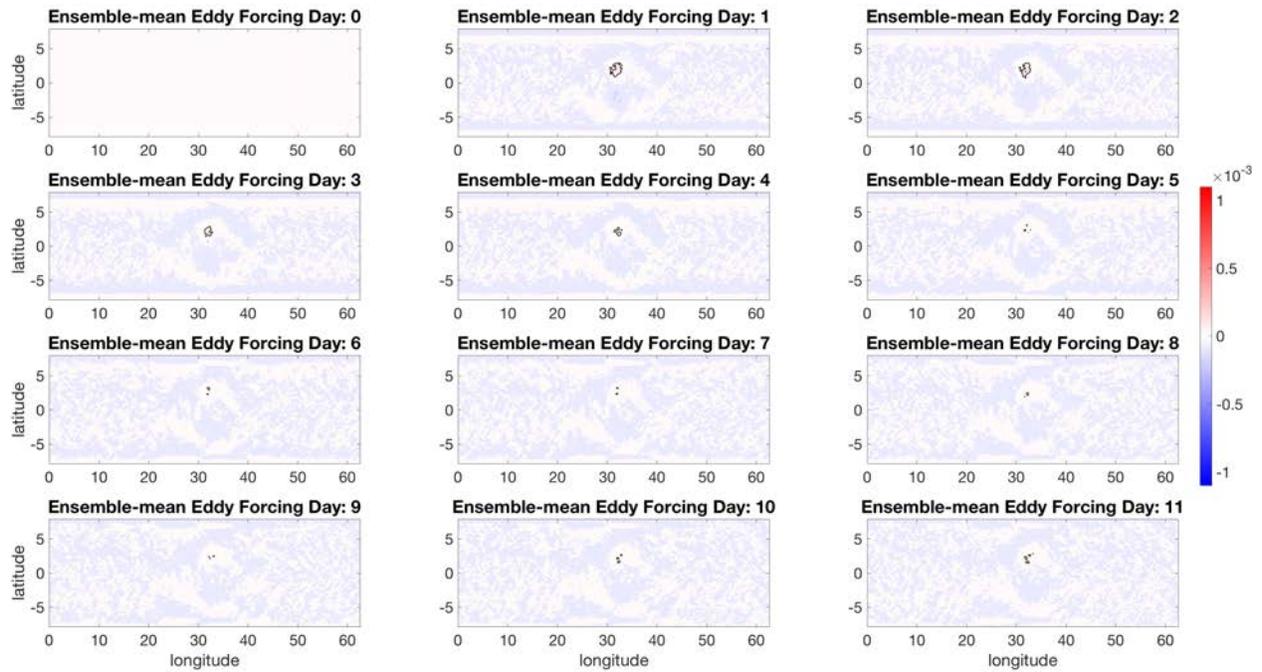

**Figure 4**. The ensemble-mean eddy PV flux divergence $-\nabla \cdot [\mathbf{v}'q']$ in a large-ensemble of a two-layer QG model. Note the same contour levels to that in Fig. 3.

# 3 Re-examining the results in Shutts (1983)

The fully nonlinear result in the last section is in contradiction to the positive eddy feedback hypothesis proposed by S83. The main supporting evidence in S83 was a weakly nonlinear barotropic model with a localized wavemaker. In this section, following the methodology of S83, we perform a weakly nonlinear calculation of the barotropic component of the two-layer QG model to seek the second-order induced flow. See Appendix B for details of the model. The zero-th order flow is based on the mature stage of the composite barotropic PV $q_{block}$ from the two-layer model.

In Case 1, as shown in Figs. 5(a)(c), when the wave maker is placed immediately upstream of the block, the second-order induced flow shows a pattern consistent with the blocking dipole suggesting that reinforcement is possible, which is qualitatively consistent with S83.

However, in Case 2, as shown in Figs. 5(b)(d), when the wavemaker is placed far upstream, the second-order induced flow shows an opposite pattern over the block region suggesting that reinforcement is not identifiable. This is in contraction with S83, and demonstrates a strong sensitivity of the results to the location of the wavemaker. Such sensitivity to the location of the wave maker casts doubt on the interpretation of the results in S83. In fact, a similar sensitivity has been reported earlier for a barotropic model with realistic flow pattern (Maeda et al., 2000).

As hypothesized by S83, as a result of enhanced enstrophy production at the point of jet splitting, one may expect an enhanced diffusion to induce a second-order flow, whose spatial structure, according the calculations done in S83, is to reinforce the block. Note that such enhanced diffusion would exist even without a block as the eddies disperse zonally and meridionally, as long as there is a background PV gradient (such as the planetery vorticity gradient), see the results in Case 3 in Figs. 5(e)(g). Therefore, with the approach in S83, a technical challenge is to exclude the influence from the wave maker itself, which isn't straightforward for a barotropic model that has to place a wave maker somewhere.

To eliminate the influence from the location of the wave maker, we conduct an ensemble of weakly nonlinear calculations. For each member, we place the wave maker at a different longitude. The difference in the longitude between adjacent members is $1.25\pi L_d$, where $L_d$ denotes the radius of deformation in the QG model. With 16 members, the ensemble covers the entire zonal extent. Because of cancelations between simulations with wave makers close to the block and far upstream of the block, the ensemble-mean second-order induced flow (Figs. 5(f)(h)) is one order smaller than that in Cases 1 and 2. The ensemble-mean second-order induced flow shows a quadruple structure with little projection onto the blocking dipole.

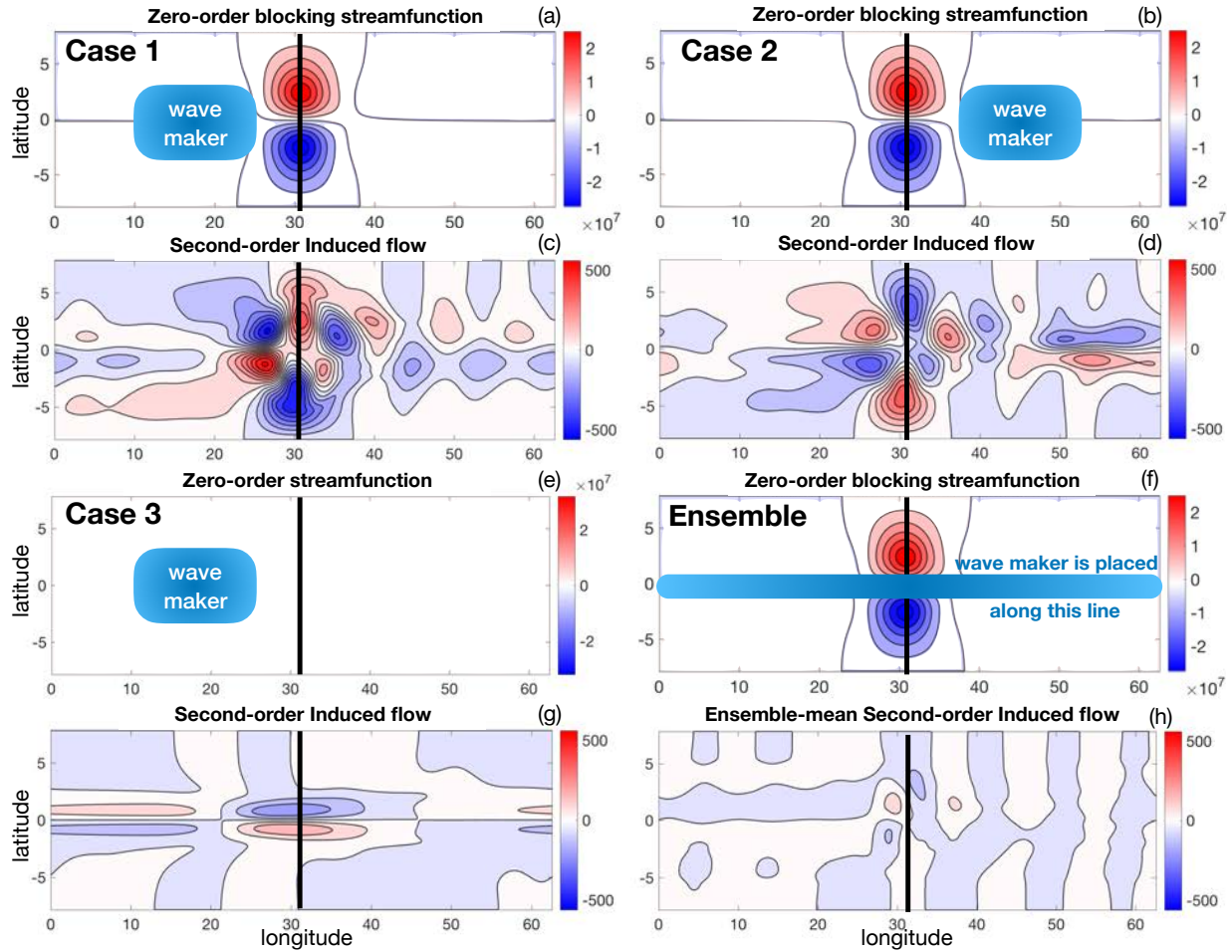

**Figure 5** (a)(c) The zero-order blocking streamfunction and the second-order induced flow for Case 1 where the wave maker is close upstream of the block. (b)(d) Similar to Case 1, but the wave maker is placed far upstream of the block. (e)(g) Similar to Case 1, but the zero-order streamfunction does not contain a block. (f)(h) Similar to Case 1, but is a 16-member ensemble average where for each member the wave maker is placed at a different longitude along the blue line marked in (f). In (a)(b)(e)(f), the contour interval for the zero-order streamfunction is $4\times10^6$ m$^2$ s$^{-1}$ and in (c)(d)(g)(h) the contour interval for the second-order induced flow is 7 m$^2$ s$^{-1}$.

## 4 Discussions

The discrepancy between the ensemble of the weakly nonlinear calculations in Section 3 and the fully nonlinear simulations in Section 2 presumably arises from the several assumptions made in the weakly nonlinear calculations in S83. These assumptions include:

(1). Finite-amplitude synoptic eddies can be qualitatively treated as small-amplitude waves;

(2). The baroclinic-barotropic interaction is not essential;

(3). In the presence of a block, the horizontal distribution of the synoptic eddy activity can be simplified as a localized wavemaker.

Beyond these assumptions made, why is the enhanced diffusion not working as expected in S83 to provide the eddy forcing that reinforces the block? It is true that a high enstrophy production must lead to an enhanced diffusion as argued in S83, but the location of such enhancement does not coincide with the jet split region. Note that wavemaker *per se* generates enstrophy. In fact, in S83's Fig. 4(d), when the wavemaker is placed away from the jet splitting region, no enhanced enstrophy can be identified at the jet splitting region. Instead, two local maxima occur near the two jet branches of the block. Lastly, why does the zonal convergence of wave activity not maintain the mature block? We speculate that as synoptic eddies strain, they can propagate along the two jet branches and disperse away meridionally, preventing the local wave activity associated with the strained eddies from building-up. These processes will be investigated in future work.

## 5 Conclusions

Despite the evidence that a sequence of events, including eddy straining, meridional shifts, and wave breaking, are observed to be concurrent with the mature stage of blocks, we show that synoptic eddies do not reinforce the block as a general positive feedback mechanism when the initial conditions are randomly sampled.

Our results illustrate the important role of the initial conditions of the pre-existing eddies in giving rise to the lifecycle of blocks. Specific configurations of pre-existing eddies must be instrumental for blocking maintenance, which, when satisfied, provide conducive condition for meridional displacement as well as the subsequent wave breaking to occur, which is the reason of the concurrency in the observations. Detailed mechanisms of how these specific initial conditions could lead to block maintenance warrant further studies.

**Appendix A: A two-layer quasi-geostrophic model**

Following Wang and Lee (2016) (WL16), the two-layer QG model with an unequal layer thickness on a beta-plane:

$$\frac{\partial q_1}{\partial t} + J(\psi_1, q_1) = -\tau^{-1}\frac{\psi_2 - \psi_1 + \psi_R}{2(2-\delta)} - \kappa\nabla^6\psi_1,$$
$$\frac{\partial q_2}{\partial t} + J(\psi_2, q_2) = \tau^{-1}\frac{\psi_2 - \psi_1 + \psi_R}{2\delta} - \gamma^{-1}\nabla^2\psi_2 - \kappa\nabla^6\psi_2, \quad (A1)$$

where the quasi-geostrophic potential vorticity (PV) is:

$$q_1 = \beta y + \nabla^2\psi_1 + \frac{\psi_2 - \psi_1}{2(2-\delta)},$$
$$q_2 = \beta y + \nabla^2\psi_2 - \frac{\psi_2 - \psi_1}{2\delta}. \quad (A2)$$

The subscripts 1 and 2 refer to the upper and lower layers, respectively.

The nondimensional $\beta$ measures the ratio of planetary vorticity gradient to vertical shear contribution. $\delta$ denotes nondimensional thickness of the lower layer at rest. The velocity field is determined by the relation $(u_i, v_i) = (-\partial\psi_i/\partial y, \partial\psi_i/\partial x)$. The time is non-dimensionalized by $L_d/U$, where horizontal length scale is $L_d = 750\, km$ and velocity scale is $U = 45\, ms^{-1}$. Hyper-viscosity is included in both layers to remove enstrophy at small scales. Ekman damping with a damping time scale $\gamma$ of 1.5 day is included in the lower layer only, and thermal relaxation of the upper layer zonal mean flow to a prescribed jet-like "radiative equilibrium state" $U_e \equiv -\partial\psi_e/\partial y = \text{sech}^2(\frac{y}{2})$ and the lower layer to zero wind is adopted with a relaxation time scale of 30 days.

The equation (A1) is solved numerically with Fourier spectral decomposition in the zonal direction and sine function decomposition in the meridional direction. As in WL16, in this study we choose a layer thickness ratio of $\delta = 0.25$. The key findings in this study have been tested with other values (e.g. standard equal-layer thickness) and have been confirmed to be insensitive to the particular value of choice. The non-dimensional channel length and width are set to $L_x = 20\pi$ and $L_y = 5\pi$ respectively. The width chosen is sufficiently large so that eddy amplitude is eligible near the walls. A sponge layer is added at both northern and southern boundaries to avoid reflecting waves.

# Appendix B: A weakly nonlinear barotropic quasi-geostrophic model

Following S83, we construct a similar weakly nonlinear barotropic QG model that is consistent with the above two-layer QG model.

Zero-order equation: the time-invariant mature block

$$U_b \frac{\partial q_{b0}}{\partial x} + J(\psi_{b0}, q_{b0}) + (\beta - \frac{\partial^2 U_b}{\partial y^2}) \frac{\partial \psi_{b0}}{\partial x} = -D(\psi_{b0}) \tag{B1}$$

First-order equation: incoming eddies and straining effects

$$(\frac{\partial}{\partial t} + U_b \frac{\partial}{\partial x}) q_{b1} + J(\psi_{b1}, q_{b0}) + J(\psi_{b0}, q_{b1}) + (\beta - \frac{\partial^2 U_b}{\partial y^2}) \frac{\partial \psi_{b1}}{\partial x} = F_1 - D(\psi_{b1}) \tag{B2}$$

Second-order equation: induced circulation

$$(\frac{\partial}{\partial t} + U_b \frac{\partial}{\partial x}) q_{b2} + J(\psi_{b0}, q_{b2}) + J(\psi_{b2}, q_{b0}) + (\beta - \frac{\partial^2 U_b}{\partial y^2}) \frac{\partial \psi_{b2}}{\partial x} = -J(\psi_{b1}, q_{b1}) - D(\psi_{b2}) \tag{B3}$$

where $U_b = \frac{2-\delta}{2} U_1 + \frac{\delta}{2} U_2$ denotes barotropic density-weighted zonal-mean zonal velocity. $q_{b0}, q_{b1}, q_{b2}$ denote zero-th order, first order, and second order of the barotropic PV with $q_b = \frac{2-\delta}{2} q_1 + \frac{\delta}{2} q_2$, and similar notation is used for streamfunction $\psi_{b0}, \psi_{b1}, \psi_{b2}$.

A wave maker is prescribed by:

$$F_1 = 1.0^{-12} \cdot \sin\{\frac{\pi(x-x_0)}{\Delta x}\} \cos\{\frac{3\pi(x-x_0-t)}{\Delta x}\} \exp\{-4(\frac{y-y_0}{\Delta y})^2\} \tag{B4}$$

where $x_0$ is the starting longitude of the wave maker, $y_0$ is the mid-point of the channel, $\Delta x = L_x / 4$ and $\Delta y = L_y / 4$ are the zonal and meridional extent of the wave maker.

## Acknowledgments

The authors acknowledge the computational resources at NCAR Yellowstone and Harvard Odyssey cluster. The authors acknowledge discussions with Dehai Luo, Akira Yamazaki, Pedram Hassanzadeh, and Brian Hoskins. This research was supported by NASA grant 80NSSC17K0267, NSF grant AGS-1552385, and a grant from the Harvard Global Institute.